# Unveiling the 5$f$ electron hybridization process in UPd$_2$Al$_3$ via ARPES and Time-resolved PES

Jiao-Jiao Song,[1] Qi-Yi Wu,[1] Chen Zhang,[1] Steve M. Gilbertson,[2] Peter S. Riseborough,[3] Ján Rusz,[4] John J. Joyce,[2] Kevin S. Graham,[2] Clifford G. Olson,[5] Paul H. Tobash,[2] Eric D. Bauer,[2] Bo Chen,[1] Hao Liu,[1] Yu-Xia Duan,[1] Peter M. Oppeneer,[4] George Rodriguez,[2] Tomasz Durakiewicz,[6] and Jian-Qiao Meng[1, *]

[1]*School of Physics, Central South University, Changsha 410083, Hunan, China*
[2]*Los Alamos National Laboratory, Los Alamos, New Mexico 87545, USA*
[3]*Temple University, Philadelphia, Pennsylvania 19122, USA*
[4]*Department of Physics and Astronomy, Uppsala University, Box 516, S-75120 Uppsala, Sweden*
[5]*Ames Laboratory, Iowa State University, Ames, Iowa 50011, USA*
[6]*Idaho National Laboratory, Idaho Falls, ID 83415 USA*
(Dated: Tuesday 10$^\text{th}$ September, 2024)

This study investigates the 5$f$-electron–conduction electron hybridization process in the heavy fermion superconductor UPd$_2$Al$_3$ using a combination of angle-resolved photoemission spectroscopy (ARPES) and time-resolved photoemission spectroscopy (tr-PES). ARPES measurements reveal the formation of a hybridization gap at a temperature of approximately 75 K, which becomes more pronounced as the temperature decreases. Notably, the persistence of a flat U 5$f$ band at temperatures well above the hybridization onset challenges conventional understanding. Our findings demonstrate a non-monotonic temperature dependence of the quasiparticle relaxation time, with an anomalous decrease at 20 K, suggesting complex electronic and magnetic interactions. These findings provide detailed insights into the 5$f$-electron hybridization process in UPd$_2$Al$_3$, with significant implications for the understanding of heavy fermion superconductivity and the role of 5$f$-electron hybridization in uranium-based materials.

PACS numbers: 74.25.Jb,71.18.+y,79.60.-i

Heavy fermion (HF) systems are of significant interest in condensed matter physics due to their rich electronic behavior, arising from the intricate interplay between localized $f$-electrons and itinerant conduction electrons [1, 2]. These systems serve as a crucial platform for exploring quantum criticality and the emergence of unconventional superconductivity, which deviate from the traditional Bardeen-Cooper-Schrieffer (BCS) theory. The ability of heavy fermion materials to exhibit both magnetic order and superconductivity, often in close proximity to a quantum critical point, makes them a fascinating subject of study for understanding complex quantum phenomena. This complexity underscores the importance of understanding the behavior of $f$-electrons, which is governed by the competition between Ruderman-Kittel-Kasuya-Yosida (RKKY) and Kondo interactions [3].

In HF superconductors, the hybridization between localized $f$-electrons and conduction electrons, driven by the Kondo interaction, leads to the formation of a hybridization gap at the Fermi energy ($E_F$) [4–6]. This gap is a key determinant of the electronic properties in HF materials, influencing not only their low-temperature magnetic properties but also the emergence of unconventional superconductivity. Understanding this hybridization process is vital for unraveling the complex mechanisms that govern the behavior of these systems, especially near quantum critical points.

Probing the behavior of $f$-electrons and the evolution of the hybridization gap with high precision remains challenging, largely due to the limitations of traditional experimental techniques. Advanced methods such as angle-resolved photoemission spectroscopy (ARPES) [7–9] and time-resolved photoemission spectroscopy (tr-PES) [10, 11] offer critical advantages, providing direct insights into the electronic structure and dynamics with high energy and temporal resolution. These techniques are particularly suited to capturing the subtle changes in hybridization and the associated quasiparticle dynamics, which are crucial for understanding the physics of heavy fermion systems.

While previous studies have hinted at the existence of a hybridization gap in UPd$_2$Al$_3$, a comprehensive understanding of its temperature-dependent evolution and the interplay with other low-temperature phenomena remains elusive. Our combined ARPES and tr-PES approach aims to fill this gap by employing ARPES and tr-PES to systematically investigate the hybridization process across a range of temperatures in UPd$_2$Al$_3$. By doing so, we seek to provide deeper insights into the mechanisms underlying the material's antiferromagnetic (AFM) and superconducting (SC) states.

UPd$_2$Al$_3$, known for its AFM phase transition at 14 K ($T_N$) and subsequent unconventional SC state at 2 K ($T_c$) [12–15], is an ideal candidate for this study. The large observed magnetic moment of 0.85 $\mu_B$ suggests significant localization of the $f$ electrons [16, 17]. However, the substantial jump in specific heat at $T_c$ indicates the presence of moderately heavy fermion behavior, which requires delocalized $f$ electrons and implies their involvement in the superconducting transition [15]. The interplay between the AFM and SC states has been suggested from inelastic neutron-scattering studies [18, 19]. Previous studies have reported that the hybridization gap in UPd$_2$Al$_3$ opens at $T^* \sim 65$ K [20, 21], with optical conductivity measurements identifying a gap around 10 meV below $\sim 50$ K [22]. Quasiparticle scattering spectroscopy further reveals that the hybridization gap remains finite up to $\sim 28$ K, slightly above $T_N$ [23].

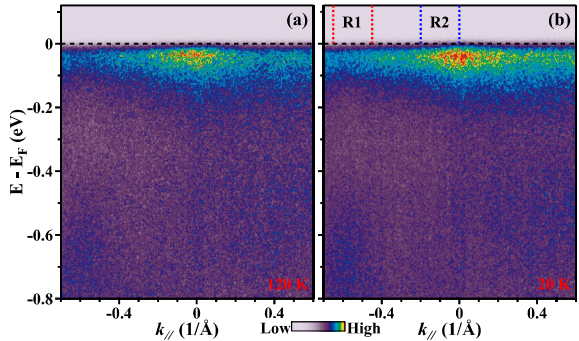

FIG. 1. ARPES of UPd$_2$Al$_3$. (a) and (b) ARPES images measured with 34 eV photons, at 20 K and 120 K, respectively. The dotted red and blue blocks mark the integrated EDC windows, labeled R1 and R2, respectively.

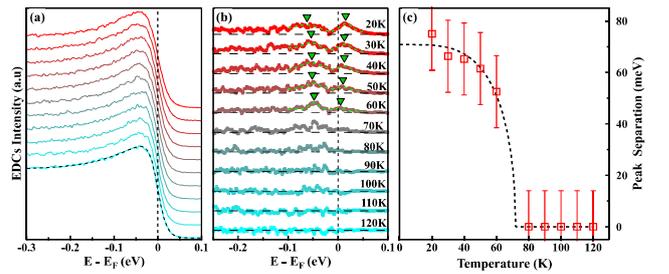

FIG. 2. Temperature dependence of electronic structure in UPd$_2$Al$_3$. (a) Temperature dependence of the EDC integrated over region R1 [Fig. 1(b)] measured with 34 eV photons. The dashed black line represents a fit to the EDC at 120 K, as detailed in the main text. (b) Difference spectra obtained by subtracting the fitted curve of the 120 K EDC from all EDCs. Green dashed lines indicate fits to these difference spectra. Triangular symbols mark the positions of the two peaks in each difference spectrum, as determined by the fitting procedure. (c) Peak separation extracted from the fits in panel (b) plotted as a function of temperature. The dashed black line represents a fit using Eq. (1).

These characteristics make UPd$_2$Al$_3$ a prototypical system for exploring the intricate dynamics of $f$-electron hybridization. By leveraging the capabilities of ARPES and tr-PES, we aim to uncover the subtle electronic changes that drive its unique low-temperature properties. The insights gained from this study on UPd$_2$Al$_3$ will not only advance our understanding of hybridization processes in HF systems but may also provide a framework for exploring similar phenomena in other correlated electron materials.

High-quality single crystals of UPd$_2$Al$_3$ were grown by the Czochralski method. High-resolution ARPES measurements were conducted at the Synchrotron Radiation Center, Stoughton, WI, using an SES 4000 hemispherical electron energy analyzer. All measurements were conducted approximately along the $\Gamma'$-$K'$ direction. Temperature-dependent ARPES measurements were performed with a photo energy of $h\nu = 34$ eV and a total energy resolution of $\sim 15$ meV. The angular resolution for all ARPES measurements was 0.20°. tr-PES measurements were employed to track the ultrafast quasiparticle (QP) dynamics using driven high-harmonic generation (HHG) and a hemispherical electron analyzer [10, 24, 25]. Samples were pumped with 1.55 eV (800 nm) laser pulses at a repetition rate of 10 kHz, with each pulse lasting 30 fs. The probing was carried out with 15 fs extreme-ultraviolet (EUV) pulses at an energy of 32.55 eV [10, 24, 25]. The effective time and energy resolutions for the tr-PES measurements were about 35 fs and 200 meV [24], respectively.

ARPES measurements (Fig. 1) reveal the presence of flat U 5$f$ bands near the $E_F$ in UPd$_2$Al$_3$, suggesting a strong photoionization cross-section for the U 5$f$ states even with 34 eV photons. These flat bands are observable at relatively high temperatures (120 K) [Fig. 1(a)], significantly higher than the onset temperature ($T^*$) of the hybridization gap [21, 22], consistent with our previous on-resonance measurements [26]. This observation aligns with those in Ce- [7, 8, 27–29], Yb- [30], and U-based [31, 32] HF materials. However, the ARPES spectra at much lower temperatures [20 K, Fig. 1(b)] show only a slight increase in spectral weight compared to higher temperatures, making direct quantification of the hybridization gap's temperature dependence challenging.

To address this, we analyzed the temperature-dependent spectral weight within region R1 [Fig. 1(b)] using energy distribution curves (EDCs) taken at temperatures ranging from 20 K to 120 K with 34 eV photons [Fig. 2(a)]. The U 5$f$ state persists at high temperatures (120 K) with minimal spectral weight change upon cooling. To isolate the contribution of hybridization, the highest temperature (120 K) EDC was fitted to a Lorentzian plus a Shirley background, multiplied by a Fermi-Dirac function, and finally convolved with the instrumental resolution [black dashed line, Fig. 2(a)]. Subtracting this fitted curve from all EDCs effectively isolated the spectral weight changes due to hybridization [Fig. 2(b)].

The difference spectra reveal the emergence of two distinct peaks below 70 K, indicating the opening of a hybridization gap. The peak separation and intensity progressively increase with decreasing temperature. Figure 2(c) shows the temperature dependence of the peak separation extracted from the fits in Fig. 2(b). The separation emerges around 70 K and increases as the tem-perature decreases. While not directly equivalent to the hybridization gap size, its evolution reflects the temperature dependence of the gap itself. The behavior of the peak separation can be described by an empirical mean-field equation proposed by Gross et al. [33] to capture the temperature dependence of the hybridization process:

$$\Delta(T) = \Delta_0 \tanh\left[\frac{\pi T_o}{\Delta_0}\sqrt{\alpha\left(\frac{T_o}{T}-1\right)}\right]\Theta(T_o - T), \quad (1)$$

where $\Delta_0$ is the zero-temperature separation, $\alpha$ is a parameter related to the specific pairing state, $T_o$ is the hybridization gap opening temperature, and $\Theta$ is the Heaviside step function. This fitting approach, shown by the dashed black line in Fig. 2(c), yields a gap opening temperature of $\sim 73 \pm 10$ K and a zero-temperature separation of $\Delta_0 \sim 71$ meV. Similar results were obtained from the analysis of EDCs in region R2 in Fig. 1(b) [see Fig. S2 in Supplemental Material [34] for details].

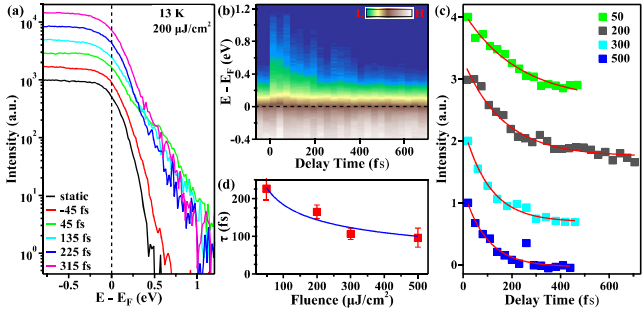
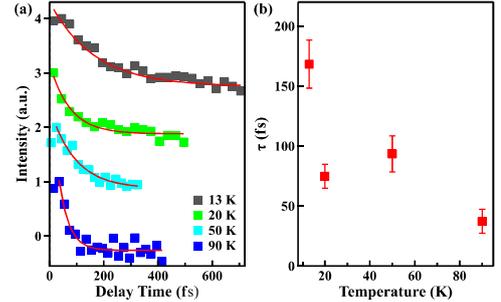

FIG. 3. Fluence dependence of ultrafast electronic dynamics in UPd$_2$Al$_3$ at 13 K. (a) Transient photoemission spectra measured at a pump fluence of 200 $\mu$J/cm$^2$. The curves are offset for better clarity. The black line represents the ground-state PES spectrum (static, unpumped). (b) Pseudocolor of the time-dependent photoelectron intensity at a pump fluence of 200 $\mu$J/cm$^2$. (c) Time-dependent spectral weight integrated within the energy window [0, 0.4 eV] for various pump fluences (in $\mu$J/cm$^2$). The red lines are single-exponential decay fits. (d) Extracted lifetimes ($\tau$) as a function of pump fluence. The blue line represents a fit to the data using the RT model.

FIG. 4. Temperature dependence of ultrafast electronic dynamics in UPd$_2$Al$_3$ with 200 $\mu$J/cm$^2$ pump fluence. (a) Time-dependent spectral weight integrated within the energy window [0, 0.4 eV] for various temperatures. Red lines are single-exponential decay fits. (b) Extracted lifetimes ($\tau$) versus temperature.

A narrow energy gap near $E_F$ is known to hinder relaxation after excitation in HF materials [35–40] and high-temperature superconductors [41–44]. Here, we employ tr-PES to probe the ultrafast quasiparticle dynamics in UPd$_2$Al$_3$. Due to low count rates, angle-integrated tr-PES measurements were performed at a probe energy of 32.55 eV, ensuring sufficient photoionization cross-sections for the U 5$f$ orbital [25]. This approach enables us to examine the impact of the hybridization gap on the relaxation process and trace the temperature evolution of the hybridization process in UPd$_2$Al$_3$.

To minimize initial perturbations to the electronic structure and focus on the intrinsic dynamics, we utilized a weak laser excitation pulse with a fluence of 200 $\mu$J/cm$^2$ at a cryogenic temperature of 13 K. Figure 3(a) shows transient EDCs for various delay times after the laser pulse. Compared to the unpumped reference spectrum (solid black line), significant changes in the density of unoccupied states are observed, indicating a dynamic evolution of the electronic structure following photoexcitation. Figure 3(b) presents a 2D pseudocolor map depicting the photoemission spectra as a function of delay time ($x$ axis) and energy ($y$ axis). A prominent feature is the emergence of a broad "excitation continuum" upon laser excitation, extending up to $\approx$ 1.2 eV above $E_F$. This continuum represents the initial electronic excitation. Notably, the intensity of this continuum decays over time, with slower decay observed closer to $E_F$ —a well-established behavior for electronic excitations on metal surfaces.

To quantify the relaxation dynamics, the spectral weight within the [0, 0.4 eV] energy window was integrated [Fig. 3(c)], along with pump fluences of 50, 300, and 500 $\mu$J/cm$^2$ [see Fig. S2 in Supplemental Material [34] for more data]. The time-dependent spectral weight for various pump fluences demonstrates a dependence of the excited electronic state's lifetime ($\tau$) on pump fluence, with higher fluences corresponding to shorter lifetimes (faster relaxation). The time-dependent spectral weight was well-fitted by a single-exponential decay function, as shown by the solid red lines. Extracted lifetimes ($\tau$) as a function of pump fluence are plotted in Fig. 3(d). The observed decrease in $\tau$ with increasing fluence aligns with expectations. The Rothwarf-Taylor model [35] fit (blue line) suggests that even at the maximum pump fluence of 500 $\mu$J/cm$^2$, the hybridization gap in UPd$_2$Al$_3$ remains open [see details in Supplemental Material [34]].

To investigate the temperature evolution of the hybridization process in UPd$_2$Al$_3$, temperature-dependent tr-PES measurements were conducted with a pump fluence of 200 $\mu$J/cm$^2$ at 13, 20, 50, and 90 K. Figure 4(a) shows the spectral weight within the [0, 0.4 eV] energy window [see Fig. S3 in Supplemental Material [34] for more data]. These measurements reveal how the hybridization gap evolves with temperature, providing in-sights into the electronic structure and correlations in UPd$_2$Al$_3$. Figure 4(b) displays the extracted quasiparticle lifetimes as a function of temperature. As expected, due to the opening of the hybridization energy gap at low temperatures, the relaxation time at low temperatures is significantly longer than at high temperatures (90 K). This indicates that the hybridization process slows down the relaxation dynamics at lower temperatures, which is consistent with the increased energy gap inhibiting the decay of excited states and longer quasiparticle lifetimes corresponding to narrowing of relevant peaks in the density of states. These findings underscore the temperature sensitivity of the hybridization gap in UPd$_2$Al$_3$ and provide valuable insights into the mechanisms governing its electronic properties. The observed changes in spectral weight suggest that the hybridization process is highly temperature-sensitive, which could have significant implications for understanding the materials heavy fermion behavior.

Notably, the relaxation time exhibits a non-monotonic temperature dependence, with an anomalous decrease observed at 20 K before increasing again. This behavior suggests a complex interplay between the hybridization gap and other low-temperature electronic or magnetic phenomena in UPd$_2$Al$_3$. Two potential explanations can be considered: (i) Competition between hybridization and magnetic fluctuations: The proximity to the AFM transition at 14 K may lead to magnetic fluctuations that disrupt the coherence of the $f$-electrons. In HF systems, this disruption could weaken the hybridization process, leading to observable changes in electronic properties, such as the decrease in relaxation time at 20 K. The competition between Kondo screening, which promotes hybridization, and the RKKY interaction, which drives magnetic ordering, could be at the heart of this behavior. This suggests that UPd$_2$Al$_3$ is near a delicate balance point between these competing interactions, which could play a significant role in its unconventional superconductivity. As the temperature decreases, these fluctuations could temporarily weaken the hybridization, resulting in a faster relaxation time at 20 K, despite the overall trend of longer lifetimes at lower temperatures due to the hybridization gap. As the temperature drops further below $T_N$, the AFM phase may induce additional gapping in the electronic spectra, leading to a subsequent increase in the relaxation time. (ii) Enhanced electron-electron or electron-phonon interactions: At specific temperatures, such as 20 K, these interactions may intensify, leading to increased scattering processes and a faster relaxation time, even in the presence of the hybridization gap. These temperature-dependent interactions suggest that UPd$_2$Al$_3$ may host additional low-energy excitations or coupling mechanisms that become particularly active at certain temperatures. Such behaviors are reminiscent of other heavy fermion compounds [38, 39, 45], where the coupling of quasiparticles with some bosonic excitations has been observed to significantly influence the quasiparticle dynamics. Understanding these interactions is crucial for a more comprehensive picture of the hybridization process and its role in the material's superconducting and magnetic properties. Future investigation using techniques like neutron scattering [18, 19] is necessary to elucidate the underlying mechanisms contributing to this non-monotonic temperature dependence of the relaxation dynamics.

In summary, we have investigated the hybridization process in antiferromagnetic heavy fermion superconductor UPd$_2$Al$_3$ using ARPES and tr-PES. The persistence of a flat U 5$f$ band at high temperatures and the non-monotonic temperature dependence of the relaxation time provide new insights into the mechanisms of heavy fermion physics. These findings not only advance our understanding of the complex interplay between $f$-electrons and the hybridization process, but also provide valuable insights into the emergence of unconventional superconductivity in UPd$_2$Al$_3$.

This work was supported by the National Key Research and Development Program of China (Grant No. 2022YFA1604204), National Natural Science Foundation of China (Grant No. 12074436), and the Science and Technology Innovation Program of Hunan Province (Grant No. 2022RC3068). Work at Los Alamos was performed under the U.S. Department of Energy, Office of Basic Energy Sciences, Quantum Fluctuations in Narrow Band Systems program. J. R. and P. M. O. acknowledge funding from the K. and A. Wallenberg Foundation (Grants No. 2022.0079 and No. 2023.0336) and from the Swedish Research Council (VR) (Grant No. 2022-06725).